\documentclass{PoS}

\title{$q\bar q$-potential: a critical reappraisal}

\ShortTitle{$q\bar q$-potential}

\author{\speaker{G.C. Rossi}
\\
        {Dipartimento di Fisica, Universit\`a di Roma {\it Tor Vergata}\\ \small and INFN, Sezione di Roma 2 \\ \small Via della Ricerca Scientifica, 00133 Roma, Italy}\\
        E-mail: \email{rossig\@roma2.infn.it}}

\author{M. Testa\\
       {Dipartimento di Fisica, Universit\`a di Roma {\it La Sapienza}\\ \small and INFN, Sezione di Roma 1 \\ \small P.le A.\ Moro 5, 00185 Roma, Italy}
\\
      E-mail: \email{massimo.testa@roma1.infn.it}}

\abstract{We show how it is possible to define and compute the potential between $q$ and $\bar q$ external sources in the singlet and octet (adjoint) representation of the colour group.}

\FullConference{31st International Symposium on Lattice Field Theory - LATTICE 2013\\
		July 29 - August 3, 2013\\
		Mainz, Germany}

\begin{document}

\section{Introduction}
\label{sec:INTRO}

In this talk we want to show that it is perfectly well possible to define the potential of a $q\bar q$ pair of colour sources in the adjoint representation~\cite{Brown:1979ya,McLerran:1981pb,Borgs:1983yk,Nadkarni:1986as,Bodwin:1994jh,Necco:2001gh,Kaczmarek:2002mc,Bali:2003jq,Shuryak:2004tx}, despite claims in the literature~\cite{Philipsen:2002az,Jahn:2004qr,Philipsen:2013ysa}. The construction we are going to present is unique and natural. It is based on the use of the formulation of the temporal gauge developed in refs.~\cite{Rossi:1979jf,Rossi:1980pg,Rossi:1982ag} and it is expected to be valid even beyond perturbation theory (PT). The content of this talk is mainly based on the results of ref.~\cite{Rossi:2013qba}.

\section{External sources in the temporal gauge}
\label{sec:EXSOU}

The Feynman kernel in the presence of $q \bar q$-sources in the $A_0=0$ (temporal) gauge reads 
\begin{eqnarray}
\hspace{-.8cm}&&K({\bf A_2},s_2,r_2;{\bf A_1},s_1,r_1;T)=\int_{{\cal G}_0}\!\!{\cal D}\mu(h) \,[U_h({\bf x}_{q})]_{s_2s_1} [U_h({\bf x}_{\bar q})]_{r_2r_1}^*\widetilde K({\bf A_2}^{U_h},{\bf A_1};T)\, ,\label{K12}\\ \hspace{-.8cm}&&\widetilde K({\bf A}_2,{\bf A}_1;T)=\int^{{\bf A}({\bf x},T_2)={\bf A}_2({\bf x})}_{{\bf A}({\bf x},T_1)={\bf A}_1({\bf x})} {\cal D}{\bf A}\exp{[-S_{YM}({\bf A}, A_0=0)]}\, ,\label{KTILDE12}
\end{eqnarray}
where ${\cal D}\mu(h)$ is the invariant Haar measure over the group, ${\cal G}_0$, of the (topologically trivial) time-independent gauge transformations that tend to the identity at spatial infinity. The quantity $\widetilde K({\bf A}_2,{\bf A}_1;T)$ is the Feynman kernel with assigned boundary gauge field values. Its integrand is (minus) the exponential of the Yang--Mills (YM) action in which the temporal component of the gauge field is set equal to zero ($A_0=0$), and ${\cal D}{\bf A}=\prod_{{\bf x}, T_1<t<T_2}d{\bf A}({\bf x},t)$. The ${\cal G}_0$-integration with the insertion of $U_h({\bf x}_{q})=\exp{[i\lambda^ah^a({\bf x}_q)}]$ and $U_h({\bf x}_{\bar q})^*=\exp{[-i(\lambda^a)^* h^a({\bf x}_{\bar q})}]$ group elements at the location of the $q$ and $\bar q$ sources, respectively, has the effect of projecting the ``master'' kernel, $\widetilde K$, over the desired source sector~\footnote{Actually the formalism developed in refs.~\cite{Rossi:1979jf,Rossi:1980pg} allows to discuss the case of an arbitrary number of external sources in any colour representation, generically  distributed in space.}. 

$K({\bf A_2},s_2,r_2;{\bf A_1},s_1,r_1;T)$ has an identical expression in the $A_0=0$ and Coulomb gauges~\cite{Rossi:1979jf} where a positive Hamiltonian exists (canonical gauges). States entering the spectral decomposition 
\begin{eqnarray}
\hspace{-.7cm}&&K({\bf A}_2,s_2,r_2;{\bf A}_1,s_1,r_1;T)=\sum_{k}e^{-E_kT}\psi_{k}({\bf A}_2,s_2,r_2)\,\psi^\star_{k}({\bf A}_1,s_1,r_1)\label{KSPEC12}
\end{eqnarray}
are eigenstates of the Hamiltonian with eigenvalue $E_k$ 
\begin{eqnarray}
{\cal H}\psi_{k}({\bf A},s,r)=E_k\psi_{k}({\bf A},s,r)\, ,\label{H}
\end{eqnarray}
and transform covariantly under $U_w({\bf x})\in {\cal G}_0$ according to 
\begin{eqnarray}
\psi_{k}({\bf A}^{U_w},s,r)=\sum_{s',r'}\Big{[}e^{-i\lambda^a w^a({\bf x}_q)}\Big{]}_{s\,s'}\Big{[}e^{i\lambda^a w^a({\bf x}_{\bar q})}\Big{]}_{r'\,r}\psi_{k}({\bf A},s',r')\, .\label{GAUSS}
\end{eqnarray}
This formula yields the Gauss' law in the presence of $q\bar q$ external sources, as it can be seen by taking the functional derivative of both sides of the equation with respect to $w^a({\bf x})$, and then setting $w^a({\bf x})=0$ . 

Eq.~(\ref{GAUSS}) implies that the quantity $\sum_{s,r} \psi^*({\bf A},s,r) \phi ({\bf A},s,r)$ is invariant under ${\cal G}_0$-gauge transformations, hence formally 
$\int {\cal D}{\bf A}\sum_{s,r} \psi^*({\bf A},s,r) \phi ({\bf A},s,r)\sim\infty$.
As a consequence, the scalar product in the Hilbert space of energy eigenstates with gauge transformation properties~(\ref{GAUSS}) must be defined via the Faddeev--Popov  procedure by setting~\cite{Rossi:1983hr}
\begin{eqnarray}
&&\hspace{2.5cm}(\psi,\phi) = \int {\cal D}\mu_F({\bf A})\sum_{s,r} \psi^*({\bf A},s,r) \phi ({\bf A},s,r) \, ,\label{SCALAR}\\
&&{\cal D}\mu_F({\bf A})=\Delta_{F}({\bf A})\prod_{{\bf x}} \delta[F({\bf A})]\,d{\bf A}({\bf x})\, , \qquad
1=\Delta_{F}({\bf A})\int_{{\cal G}_0} {\cal D}\mu(h) \delta[F({\bf A}^{U_h})]\, .\label{DELTA}
\end{eqnarray}
The scalar product is independent of the gauge fixing functional $F({\bf A})$ (typically one takes {\small{$F({\bf A})={\nabla}{\bf A}$}}) and provides a full completion of the temporal gauge.

The colour trace of the Feynman kernel also is gauge invariant. So for the full trace one must use the measure~(\ref{DELTA}) getting ($d_k=$ degeneracy of the energy level $E_k$)
\begin{eqnarray}
\hspace{-.7cm}&&\int {\cal D}\mu_F({\bf A})\sum_{s,r}  K({\bf A},s,r;{\bf A},s,r;T)=\sum_{k} d_k e^{-E_kT}\, . \label{INV}
\end{eqnarray}
The gauge invariance of the l.h.s.\ implies that the energy levels $E_k$ are gauge invariant, hence measurable, quantities.

\section{Energy eingenstate classification}
\label{sec:ENCLAS}

The global colour invariance of the YM action implies that $\cal H$-eigenstates belong to irreducible representations (irrep's) of SU($N_c$). As we are now going to show, to be able to define what is to be called ``the potential energy of a $q\bar q$-pair in the adjoint representation'' it is necessary to have a classification of energy eigenstates in terms of the irrep's of the global symmetry group~\cite{Rossi:2013qba}. 

Global colour rotations, $V\in {\mbox{SU}}(N_c)$, are implemented on $\cal H$-eigenstates by a unitary operator, ${\cal U}(V)$, that commutes with the Feynman kernel~(\ref{K12}), $[{\cal U}(V),K]$, and acts on the states in~(\ref{H}) according to the formula (see~(\ref{GAUSS}))
\begin{equation}
{\cal U}(V) \psi({\bf A},s,r)= V_{ss'} \psi({\bf A}^V,s',r') V^\dagger_{r'r}\, .\label{TRANSF}
\end{equation} 
$\cal H$-eigenfunctionals in the $q\bar q$ sector can be parametrized in the form 
\begin{eqnarray}
\psi({\bf A},s,r) = \phi ({\bf A}) \delta_{sr} + \phi_a ({\bf A})\lambda^a_{sr} \, ,\label{PARAM}
\end{eqnarray}
with global transformation properties (eq.~(\ref{TRANSF}))
\begin{eqnarray}
{\cal U}(V) \psi({\bf A},s,r) = \phi ({\bf A}^V) \delta_{sr} + \phi_a ({\bf A}^V)V_{ss'}\lambda^a_{s'r'}V^\dagger_{r'r} \, .\label{PARTR}
\end{eqnarray}
We see that under global colour rotations the state $\psi$ gets two contributions: a ``colour-spin'' piece coming from the action of $V$ on source indices and an ``orbital'' one coming from the rotation of the gauge field, ${\bf A}\rightarrow {\bf A}^V$. This entanglement of spin and orbital indices is what makes non-trivial the identification of the adjoint $q\bar q$-potential. 

The difficulty is solved by looking at the way $\cal H$-eigenstates behave at vanishing values of the gauge field. The key observation is that $\psi({\bf A},s,r)$, as well as the colour transformed state ${\cal U}(V) \psi({\bf A},s,r)$, must span a unique irrep.\ for any value of $\bf A$, hence also at ${\bf A}={\bf 0}$. But, since at ${\bf A}={\bf 0}$ the orbital contribution is absent in eq.~(\ref{PARTR}), we conclude that in the formula 
\begin{eqnarray}
{\cal U}(V) \psi({\bf 0},s,r) = \phi ({\bf 0}) \delta_{sr} + \phi_a ({\bf 0})V_{ss'}\lambda^a_{s'r'}V^\dagger_{r'r}\, , \label{PARTR0}
\end{eqnarray}
the two terms in the r.h.s.\ cannot be simultaneously non-vanishing, otherwise the state ${\cal U}(V) \psi({\bf 0},s,r)$ would belong to the reducible $[S] \oplus [N_c^2 - 1]$ representation. Thus we have the following three possibilities 

\qquad $\bullet$ $\phi_a ({\bf 0}) =0$ and  $\phi ({\bf 0}) \neq 0$

\qquad $\bullet$  $\phi ({\bf 0}) = 0$ and $\phi_a ({\bf 0}) \neq 0$ (for some $a$) 

\qquad $\bullet$ $\phi ({\bf 0}) = \phi_a ({\bf 0}) =0$ \\
in correspondence to different types of irrep's discussed in detail in ref.~\cite{Rossi:2013qba}. The result of the analysis is that $\cal H$-eingenstates can be classified in four classes according to how spin and of orbital functionals transform under global colour rotations. More precisely one gets 

(1)\,\,spin singlet $\otimes$ orbital singlet states ($\phi_a ({\bf 0}) = 0$ \& ${\phi ({\bf 0}) \neq 0}$)
\begin{eqnarray}
&&\psi^{[S]}_{[S]} ({\bf A})= \phi ({\bf A}) I\, , 
\quad{\mbox{with}}\quad \phi ({\bf A}^V) = \phi ({\bf A})\in [S]_{\rm orbit}\nonumber
\end{eqnarray}

(2)\,\,spin adjoint $\otimes$ orbital singlet states\, ($\phi ({\bf 0}) = 0$ \& ${\phi_a ({\bf 0}) \neq 0}$)
\begin{eqnarray}
&&\psi_{[Ad]}^{[S]}({\bf A}) = \lambda^a \phi_a ({\bf A})\, ,
\quad{\mbox{with}}\quad \phi_a ({\bf A}^V) = \phi_a ({\bf A})\in [S]_{\rm orbit} \nonumber
\end{eqnarray}

(3)\,\,spin singlet $\otimes$ orbital $[\alpha]$ states \,($\phi^{[\alpha]}_{m} ({\bf 0}) = 0$)
\begin{eqnarray}
\hspace{-.5cm}&&\psi^{[\alpha]}_{m\,[S]}({\bf A}) = \phi^{[\alpha]}_{m} ({\bf A}) I \, ,\quad{\mbox{with}}\quad \phi^{[\alpha]}_{m}({\bf A}^V) = R^{[\alpha]}_{mm'} (V) \phi^{[\alpha]}_{m'} ({\bf A})\in [\alpha]_{\rm orbit} \nonumber
\end{eqnarray}

(4)\,\,spin adjoint $\otimes$ orbital $[\beta]$ states combined in the irrep.\ $[\alpha']$ 
($\phi_{a k} ({\bf 0}) =0$)
\begin{eqnarray}
\hspace{-.5cm}&&\psi^{[\alpha']}_{m\,[Ad]}({\bf A}) = \lambda^a \phi_{a k} ({\bf A}) \, ,\quad{\mbox{with}}\quad 
{\phi_a}_k ({\bf A}^V) = R^{[\beta]}_{kk'} (V) \phi_{a k'} ({\bf A})\in [\beta]_{\rm orbit} \nonumber
\end{eqnarray}
A look at this list shows that the interesting singlet and adjoint $q\bar q$-potentials should be identified with the energies of the lowest states contributing to channels (1) and (2), respectively. Clearly source indices arranged in an adjoint representation are also present in channel (4). But they are entangled with the indices of the  representation $[\beta]$ to which the gluons belong to make up the irrep.\ $[\alpha']$. The lowest eigenvalues among those associated to this kind of states have been sometimes proposed as possible definitions of the $q\bar q$-potential in the adjoint representation~\cite{Bali:2003jq}. 

In our opinion this is not correct, as we will discuss in the next sections where we show how to solve the problem of singling out the states (1) and (2) from all the others. For the sake of clarity it is useful to separately discuss PT and non-perturbative (lattice) approaches.

\section{Extracting singlet and adjoint potentials in perturbation theory}
\label{sec:ESAPT}

In view of the fact that at vanishing values of the boundary gauge fields only the states labeled by (1) and (2) above survive, the most obvious formula that allows identifying singlet and adjoint $q\bar q$-potentials is
\begin{eqnarray}
\hspace{-1.2cm}&&K({\bf 0},s_2,r_2;{\bf 0},s_1,r_1;T)=
\Big{|}\phi({\bf 0})\Big{|}^2\frac{1}{N_c} 
\delta_{s_2 r_2}\delta_{r_1s_1}\,e^{-E^{[S]}T}+ \sum_{a} \Big{|}\phi_a ({\bf 0})\Big{|}^2 \,\sum_b\lambda^b_{s_2r_2}\lambda^b_{r_1s_1}\,e^{-E^{[Ad]}T}\, .\label{HOM}
\end{eqnarray}
Setting ${\bf A}_1={\bf A}_2= {\bf 0}$ kills, in fact, all orbital gluon excitations. 
For this reason we actually conjecture that only two terms are present in $K({\bf 0},s_2,r_2;{\bf 0},s_1,r_1;T)$. The conjecture was verified to hold in PT~\cite{Rossi:1982ag,Leroy:1982rg,Leroy:1986uc,Leroy:1990eh}.

Alternatively we can consider the projection over the $\delta_{r_2s_2}\delta_{s_1r_1}$ and $2\sum_a\lambda^a_{r_2s_2}\lambda^a_{s_1r_1}$ tensors of the partially traced kernel, $K_{s_2r_2;s_1r_1}(T) \equiv\int {\cal D}\mu_F({\bf A}) K({\bf A},s_2,r_2;{\bf A},s_1,r_1;T)$, yielding 
\begin{eqnarray}
&&\sum_{s_2r_2s_1r_1}\!\!\frac{1}{N_c}\delta_{r_2s_2}\delta_{s_1r_1} K_{s_2r_2;s_1r_1}(T) \stackrel{T\to \infty}\rightarrow e^{-E^{[S]}T}+\ldots +D_{[\alpha]}e^{-E^{[\alpha]}T}+\ldots\, ,\label{PTKSIN}\\
&&\sum_{s_2r_2s_1r_1} 2\sum_a \lambda^a_{r_2s_2}\lambda^a_{s_1r_1} K_{s_2r_2;s_1r_1}(T) \stackrel{T\to \infty}\rightarrow (N_c^2\!-\!1)\, e^{-E^{[Ad]}T}+\ldots + D_{[\alpha']} e^{-E^{[\alpha']}T}+\ldots\, ,\label{PTKAD}
\end{eqnarray}
where the coefficients in front of the exponentials are the dimensions of the corresponding irrep's. Singlet and adjoint potentials are extracted from the leading exponentials in the expansions~(\ref{PTKSIN}) and~(\ref{PTKAD}), respectively. 

\section{Extracting singlet and adjoint potentials in numerical simulations}
\label{sec:ESANS}

Although the formulae of the previous section can be formally rewritten in the lattice language, the implementation of the computational strategies outlined above in actual lattice simulations is undermined by a serious practical problem. To see its origin and implications consider the situation in which the boundary gauge integration in~(\ref{K12}) is extended to the group $\overline{\cal G}_0={\cal G}_0\otimes{\mbox{SU}}(N_c)$ that also includes global colour rotations. In this case to the spectral representation of the colour averaged kernel
\begin{eqnarray}
\hspace{-.8cm}&&\overline{K}({\bf A_2},s_2,r_2;{\bf A_1},s_1,r_1;T)=\int_{\overline{\cal G}_0}\!\!{\cal D}\bar\mu(h) \,[\overline{U}_h({\bf x}_{q})]_{s_2s_1} [\overline{U}_h({\bf x}_{\bar q})]_{r_2r_1}^*\widetilde K({\bf A_2}^{\overline{U}_h},{\bf A_1};T)\, ,\label{OVK12}
\end{eqnarray}
only global colour singlet eigenfunctional will contribute. As a results, the states labelled by (2) in the classification given in sect.~\ref{sec:ENCLAS} disappear from the spectrum, while among the states labelled by (3) and (4) only global singlets (i.e.\ states with $[\alpha]=[S]$ and $[\alpha']=[S]$, respectively) will survive. This feature reflects itself in the structure of eqs.~(\ref{HOM}) and~(\ref{PTKAD}). In eq.~(\ref{HOM}) we will be left with only the first term, while in eq.~(\ref{PTKAD}) precisely the interesting first term, corresponding to the adjoint potential, will be missing!

The problem we have pointed out has its origin in the fact that in lattice simulations it is extremely difficult in practice to avoid integrating over global colour rotations when one tries to implement numerically the temporal gauge fixing procedure. The reason is that, when temporal links are successively transformed to unit matrix, the accumulated final gauge transformation living on the last temporal link is in no way restricted to belong to the group ${\cal G}_0$ (unless one does something special, like it was proposed in ref.~\cite{Rossi:2013qba}).

A possible way to turn this nuisance into a benefit is to exploit the orthogonality of group characters to filter out the desired global colour irrep's among those contributing to the character-weighted kernel ($\chi^{[\gamma]}(V)$ is the character of the representation $[\gamma]$)
\begin{eqnarray}
\hspace{-.8cm}{\overline K}^{[\gamma]}_{s_2,r_2;s_1,r_1}(T) \equiv\int_{{\rm {SU}}(N_c)} \!\!{\cal D} V \,(\chi^{[\gamma]}(V))^*V_{s_2s_3}V_{r_2r_3}^*  \int {\cal D} \mu_F({\bf A}) K({\bf A}^V,s_3,r_3;{\bf A},s_1,r_1;T) \, .\label{CHARK}
\end{eqnarray}
The formula~(\ref{CHARK}) with $[\gamma]=[S]$ is nothing but eq.~(\ref{OVK12}) where, after setting ${\bf A_1}={\bf A_2}={\bf A}$, one has integrated over ${\bf A}$ with the measure~(\ref{DELTA}). To show this one needs to separate out the ${\mbox{SU}}(N_c)$-integration within ${\overline{\cal G}_0}$ by writing $\overline U_h\in \overline{\cal G}_0$ as $\overline U_h=U_h V\, ,\, U_h \in{\cal G}_0$. The choice $[\gamma]=[S]$ then gives  
\begin{eqnarray}
&&\overline K_{s_2,r_2;s_1,r_1}^{[S]}(T) =\int_{{\rm SU}(N_c)} {\cal D} V\, V_{s_2s_3}V_{r_2r_3}^*  \int {\cal D} \mu_F({\bf A}) K({\bf A}^V,s_3,r_3;{\bf A},s_1,r_1;T) \label{KAV}
\end{eqnarray}
to which only global colour singlets from states of the type (1), (3) and (4) can contribute. One can now identify the singlet $q\bar q$-potential as the lowest eigenvalue in channel (1), under the (quite reasonable) assumption that all other singlet eigenvalues, and in particular those contributing to channels (3) and (4), are larger.

Similarly, if one takes the filtering character to be that of the adjoint representation, i.e.\ if one sets $\chi^{[\gamma]}(V)=\chi^{[N^2_c-1]}(V)=2\sum_a{\mbox{Tr}}[\lambda^a V\lambda^a V^\dagger]$ in eq.~(\ref{CHARK}), only the states of channels (2), (3) and (4) that belong to a global adjoint irrep.\ will be selected. One can identify the adjoint $q\bar q$-potential as the lowest eigenvalue in channel (2), under the (quite reasonable) assumption that all other adjoint eigenvalues, and in particular those contributing to channels (3) and (4), are larger.

As discussed in~\cite{Rossi:2013qba}, the computational strategy outlined above could be implemented in lattice simulations by identifying, in every generated lattice gauge configuration, the global colour rotation $V$ in~(\ref{CHARK}) with the gauge transformation, accumulated at the final time in the process of fixing the temporal gauge, that lies at very large (infinite) spatial distance from the sources. Actual simulations using this approach turned out to be rather noisy. 

Interesting numerical results confirming the general analysis illustrated in this section  (in particular the existence of the undesired colour averaging of the simulated kernel) can be found in ref.~\cite{PREL}. 

\section{Conclusions and outlooks}
\label{sec:CONCLO}

Relying on the formulation of the YM theory in the temporal gauge (but one could equally well work in the fully equivalent Coulomb gauge~\cite{Rossi:1979jf,Rossi:1980pg}), we have shown that there exists a unique (and quite natural) definition of the adjoint $q\bar q$-potential, valid even beyond PT. We have provided explicit formulae allowing the calculation of singlet and adjoint potentials from the knowledge of the Feynman propagation kernel in the presence of $q\bar q$ external colour sources. 

We have discussed the origin of certain difficulties intrinsic to lattice simulations that have until now prevented the non-perturbative calculation of the {\it bona fide} $q\bar q$ adjoint potential defined in eqs.~(\ref{HOM}) or~(\ref{PTKAD}). A strategy to overcome these difficulties is presented in sect.~\ref{sec:ESANS}. Some preliminary promising lattice results obtained exploiting the approach we have outlined there are reported in ref.~\cite{PREL}.

\acknowledgments 
We thank G. Bali, O. Philipsen, A. Pineda and M. Wagner for discussions.

\end{document}